\documentclass[twocolumn]{webofc}
\usepackage[varg]{txfonts}
\usepackage{physics,color,ulem}

\begin{document}
\title{A novel projected shell model method for nuclear level density}
\author{\firstname{Yang} \lastname{Sun}\inst{1}\fnsep\thanks{\email{sunyang@sjtu.edu.cn}} \and
        \firstname{Jiaqi} \lastname{Wang}\inst{1} \and
        \firstname{Saumi} \lastname{Dutta}\inst{1}
        \and
        \firstname{Long-Jun} \lastname{Wang}\inst{2}
        % etc.
}

\institute{School of Physics and Astronomy, Shanghai Jiao Tong University, Shanghai 200240, China
\and
    School of Physical Science and Technology, Southwest University, Chongqing 400715, China}

\abstract
{We introduce a shell model method for calculating nuclear level density (NLD) generally applicable for  arbitrarily heavy deformed nuclei. The method is novel because of our use of physical guidance for the construction of its configuration space and the computational breakthrough with the Pfaffian algorithm. Taking a well deformed even-even $^{164}$Dy nucleus as an example, we solve exactly the eigenvalue equation to obtain a large ensemble of eigenstates of angular momentum and parity. Our results indicate a potential need to revise some common understanding of the structural behavior in the pair-breaking region where the structure changes in NLD of the compound nuclei would be sensitive for cross-sections of neutron capture.}
\maketitle

\section{Introduction}\label{intro}
Nuclear level density (NLD) is one of the most essential structural ingredients in statistical compound nuclear Hauser-Feshbach reaction theory \cite{Hauser-Feshbach} and therefore, it is important in the study of nuclear astrophysics to understand various nucleosynthesis processes \cite{Rauscher}. Besides, the knowledge is compulsory in various technological applications, e.g., in reactor engineering \cite{ADS_1}, in nuclear waste management \cite{transmutation}, and in nuclear medicine sectors, e.g., to optimize the production cross sections of medical radioisotopes used for therapeutic purposes \cite{biomedic}. 

Experimental information on nuclear levels is generally far from complete, and direct measurement is available mainly for stable nuclides. For example, while discrete nuclear levels from spectroscopic measurements are limited to only low excitation energies, at the high excitation energy side, level densities for stable isotopes can be estimated only at particle separation energies, using the experimental resonance spacing data. %Progress has also been made recently to derive level densities from particle evaporation experiments \cite{particle_evaporation_1,particle_evaporation_2,evaporation_3} which also instills model-dependency as it requires the knowledge of particle transmission coefficients to be calculated from optical models. 
%Therefore, NLDs for nuclei away from the valley of stability have to be predicted from reliable theoretical models. 
Over the past two decades, the Oslo method \cite{Oslo_0} with its  extensions, namely, the $\beta$-Oslo method \cite{beta_Oslo} and inverse Oslo method \cite{inverse_Oslo} have been widely applied to extract simultaneously NLD and strength functions from gamma-coincidence experiments. However, while the technique incorporates a certain model dependency and instils considerable uncertainties in the extracted NLD values, it requires the knowledge of complete nuclear levels at low excited energy regime in order to correctly produce the slope of the NLD curve. As mentioned above, generally, discrete spectroscopic data are not always complete, and therefore, an accurate determination of the nuclear levels at low excitation energies before the onset of the exponential behavior in the NLD curve is of significant theoretical importance. %It is of pure structure quest for which the development of a novel shell model method coupled with the application of modern many-body technique is required.

The first theoretical study on NLD was due to Hans Bethe who derived a simple analytical formula from pure statistical considerations treating the nucleus as a degenerate gas of non-interacting fermions \cite{Bethe1936, Bethe1937}. Soon after empirical models 
 %e.g., back-shifted Fermi gas model \cite{BSFG}, constant temperature model 
 \cite{BSFG,constant_temp,Gilbert-Cameron} %, composite Gilbert-Cameron model \cite{Gilbert-Cameron}, etc. 
were developed based on Bethe's concept to include shell and pairing effects which were further subject to several parametrizations \cite{ld_param_1,ld_param_2,ld_param_3,ld_param_4}. Later, the microscopic combinatorial method was developed in the spirit of Hartree-Fock mean-field theory combined with the BCS pairing, and NLDs from these models were made available for nuclei across the table \cite{combinatorial_hfb_0,combinatorial_hfb_1}. Nevertheless, methods at the mean-field level do not produce nuclear states as many-body configurations. For a quantum-mechanic description of NLD, one should solve the eigenvalue equation, $\hat H\psi = E\psi$, where $\psi$ is the many-nucleon wavefunction, and obtain all energy levels in the Hilbert space. These are real nuclear levels, which are strongly correlated single-nucleon states mediated by two-body residual interactions. Diagonalization of the Hamiltonian matrix determines the role of the single-nucleon states, and hence changes in structure. This treatment is important since only with such many-nucleon wavefunctions, can the method be able to describe thermodynamic properties and phase transitions of the nuclear states that consist of NLDs \cite{Nyhus2012}. However, solving exactly such an eigenvalue equation has turned out to be an impossible task, especially for mid-mass and heavy nuclei. Different simplifications \cite{moments_0,moments_1,Lanczos,stochastic} have been suggested to avoid the computational complexity of full Hamiltonian diagonalization. The Shell Model Monte Carlo method \cite{smmc_1, smmc_2} has been extensively discussed as an alternative approach for NLD calculations over the last twenty years. 

The development of novel shell-model methods for NLD by applying modern many-body techniques is crucial. The crux is how to construct efficient configurations. There is compelling experimental evidence \cite{Yb174_qp_isomer,8_qp_isomer, Dracoulis_2016, Walker2017} indicating that excited nuclear states can be described as many-body configurations built by the combination of broken nucleon pairs or quasiparticles (qp) from different orbitals. By taking these broken-pair multi-qp states as building blocks for the shell-model basis, we have proposed a novel shell-model method for the calculation of NLD in deformed nuclei. The shell-model diagonalization with two-body residual interactions yields a large ensemble of eigenstates of angular momentum and parity. We take the well-deformed stable rare-earth nucleus, $^{164}$Dy as our first example \cite{jiaqi2023}, for which both the low-energy discrete level data and the Oslo NLD data are available and which has distinct nuclear structure features at excitation energies of astrophysical interests that makes it a suitable candidate for applying our method as a testing ground.

With a deformed quasiparticle basis defined by the Nilsson+BCS calculation, we construct multi-quasiparticle (qp) states using the neutron ($a_{\nu}^{\dagger}$) and proton ($a_{\pi}^{\dagger}$) qp creation operators associated with the deformed qp-vacuum $\ket\phi$ as follows:
\begin{equation}\label{shell_config}
\begin{aligned}
\phantom{A} &\phantom{=}\{\ket\phi, a_{\nu_{i}}^{\dagger}a_{\nu_{j}}^{\dagger}\ket\phi,
a_{{\pi}_{i}}^{\dagger}a_{{\pi}_{j}}^{\dagger}\ket\phi, a_{\nu_{i}}^{\dagger}a_{\nu_{j}}^{\dagger}a_{\pi_{k}}^{\dagger}a_{{\pi}_{l}}^{\dagger}\ket\phi, \\
&   a_{\nu_{i}}^{\dagger}a_{\nu_{j}}^{\dagger}a_{\nu_{k}}^{\dagger}a_{{\nu}_{l}}^{\dagger}\ket\phi,
  a_{\pi_{i}}^{\dagger}a_{\pi_{j}}^{\dagger}a_{\pi_{k}}^{\dagger}a_{{\pi}_{l}}^{\dagger}\ket\phi, \\
& a_{\nu_{i}}^{\dagger}a_{\nu_{j}}^{\dagger}a_{\nu_{k}}^{\dagger}a_{{\nu}_{l}}^{\dagger}a_{\nu_{m}}^{\dagger}a_{\nu_{n}}^{\dagger}\ket\phi,   a_{\pi_{i}}^{\dagger}a_{\pi_{j}}^{\dagger}a_{\pi_{k}}^{\dagger}a_{{\pi}_{l}}^{\dagger}a_{\pi_{m}}^{\dagger}a_{\pi_{n}}^{\dagger}\ket\phi, \\ 
& a_{\pi_{i}}^{\dagger}a_{\pi_{j}}^{\dagger}a_{\nu_{k}}^{\dagger}a_{{\nu}_{l}}^{\dagger}a_{\nu_{m}}^{\dagger}a_{\nu_{n}}^{\dagger}\ket\phi, a_{\nu_{i}}^{\dagger}a_{\nu_{j}}^{\dagger}a_{\pi_{k}}^{\dagger}a_{{\pi}_{l}}^{\dagger}a_{\pi_{m}}^{\dagger}a_{\pi_{n}}^{\dagger}\ket\phi, ~\dots\}.\\ 
%$a_{\pi_{i}}^{\dagger}a_{\pi_{j}}^{\dagger}a_{\nu_{k}}^{\dagger}a_{{\nu}_{l}}^{\dagger}a_{\nu_{m}}^{\dagger}a_{\nu_{n}}^{\dagger}a_{\nu_{o}}^{\dagger}a_{\nu_{p}}^{\dagger}\ket\phi,  
\end{aligned}
\end{equation}
The notations show the construction of multi-qp configurations up to the 6$^{th}$ order while the ellipsis indicates the possibility of extending the configurations to include higher order multi-qp states whenever necessary. As there are multiple harmonic-oscillator shells for both active neutrons and protons in the PSM, the indices $\nu$ and $\pi$ in Eq.~\ref{shell_config} are general. This way, a complete shell model basis can be constructed by the combination of broken pairs (quasiparticles) within the concept of Tamm-Dancoff approximation, and each of these broken-pair states, in other words, multi-qp states, is assigned with a definite parity and $K$-quantum number.
\section{Projected Shell Model: a brief outline}
The calculations have been performed by the Projected Shell Model (PSM) \cite{PSM_Hara_Sun, PSM_Sun, Sun1996PReport, PSMcode}. It is reasonable to start with a deformed basis, such as that of the Nilsson model \cite{Nilsson1969}, since it naturally incorporates correlations for deformed nuclei. 
Three major harmonic oscillator shells have been taken in the calculation for the rare earth region. To be precise, they are: N = 4, 5, 6 (N = 3, 4, 5) for neutrons (protons) for the present $^{164}$Dy nucleus. The PSM Hamiltonian consists of separable forces:
\begin{equation}\label{two-body}
\hat{H}=\hat{H_{0}}-\frac{1}{2}\chi_{QQ}\sum_\mu \hat{Q}_{2\mu}^{\dagger} \hat{Q}_{2\mu}-G_{M}\hat{P}^{\dagger}\hat{P}-G_{Q}\sum_{\mu}\hat{P}_{2\mu}^{\dagger}\hat{P}_{2\mu},
\end{equation}
where, $\hat{H_{0}}$ is the spherical single-particle term including the spin-orbit force \cite{Nil-1985}. The rest terms are quadrupole-quadrupole interaction, monopole-pairing interaction, and quadrupole-pairing interaction, respectively,  where, ``$+$" (``$-$") denotes protons (neutrons), and $N$, $Z$, and $A$ are the neutron number, proton number, and mass number, respectively. The values of the coupling constants $G_{1}$ and $G_{2}$ are adjusted to reproduce the experimental odd-even mass differences in the respective nuclear mass region. These two values are fixed as $21.24$ and $13.86$ \cite{PSM_Hara_Sun}, respectively, in the present work. The quadrupole pairing force $G_{Q}$ is proportional to $G_{M}$ by an overall factor of $0.18$. The quadrupole deformation parameter is taken as $\varepsilon_2=0.28$ for the present case. 

\begin{figure}
\centering
\includegraphics[height=2.65in,width=3.35in]{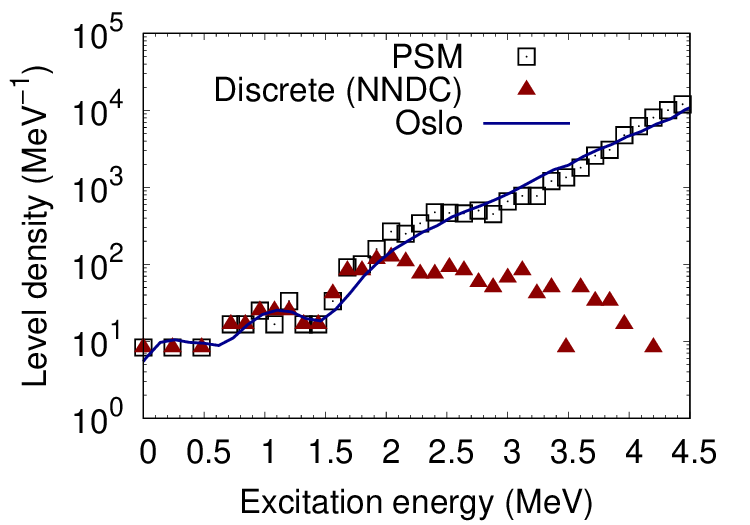}
\caption{\label{fig:NLD_compare}NLD in $^{164}$Dy obtained from PSM calculation (open squares) is compared with the Oslo data (blue solid line) \cite{Oslo1} as well as the NLD of experimentally observed discrete levels (red triangles) for which the information is taken from the website of National Nuclear Data Center \cite{NNDC_database}.}
\end{figure}
\begin{figure}
\centering
\includegraphics[height=2.75in,width=3.65in]{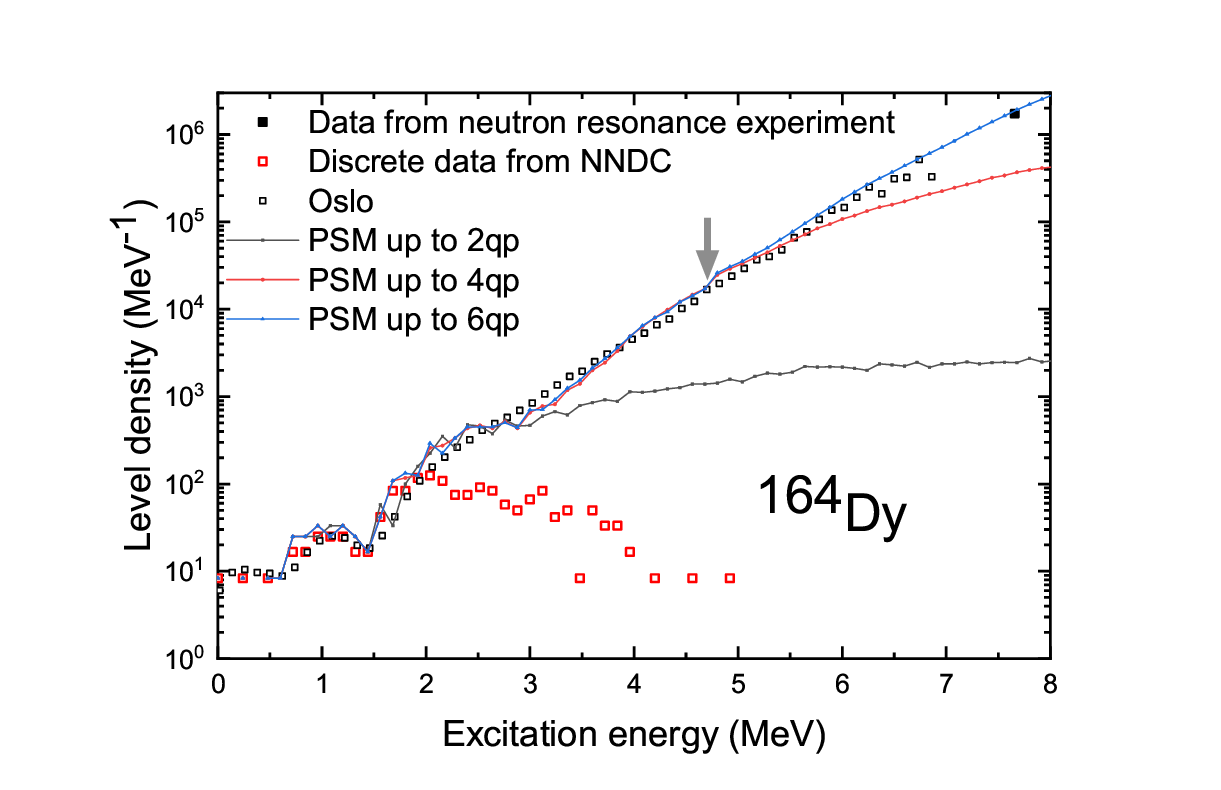}
\caption{\label{fig:dy164-piecewise-multiqp}The total level density curve in $^{164}$Dy obtained from PSM calculation (before the final step of diagonalisation but after angular momentum projection) has been shown along with the multi-qp components decomposed as 2-, 4-, and 6-qp states. Also shown are the measured discrete level density and the NLD at the neutron separation energy.}
\end{figure}
With the increasing number of multi-qp configurations for NLD calculations for heavy nuclei, a problem of combinatorial complexity emerges. To treat the complicated rotated overlap matrix elements of multi-qp states, novel computational algorithms are required \cite{Mizusaki2013}. Not long ago, the Pfaffian algorithm was derived by means of Fermion coherent states and Grassmann integral to calculate the rotated matrix elements \cite{Pfaffian_Robledo2009, Pfaffian_Bertsch_Robledo_12} and recently, this mathematical approach has been used to extend the scope of the PSM calculation to remarkably include high-order qp configurations \cite{LJWang2014} (presently, up to 10-qp states \cite{LJWang2016}) which has clearly set a breakthrough in the modern computational many-body technique and allows us to calculate spin-parity states up to $\sim$ 10 MeV of excitation in arbitrarily heavy nuclei. Angular momentum projection restores the broken rotational symmetry and transforms the configurations from the intrinsic frame to the laboratory frame. This follows the final step of diagonalisation of the two-body Hamiltonian matrix elements in the projected space, which generates the desired wavefunctions and eigenstates as a function of parity and angular momentum.
\begin{figure}
    \centering
    \includegraphics[scale=0.39]{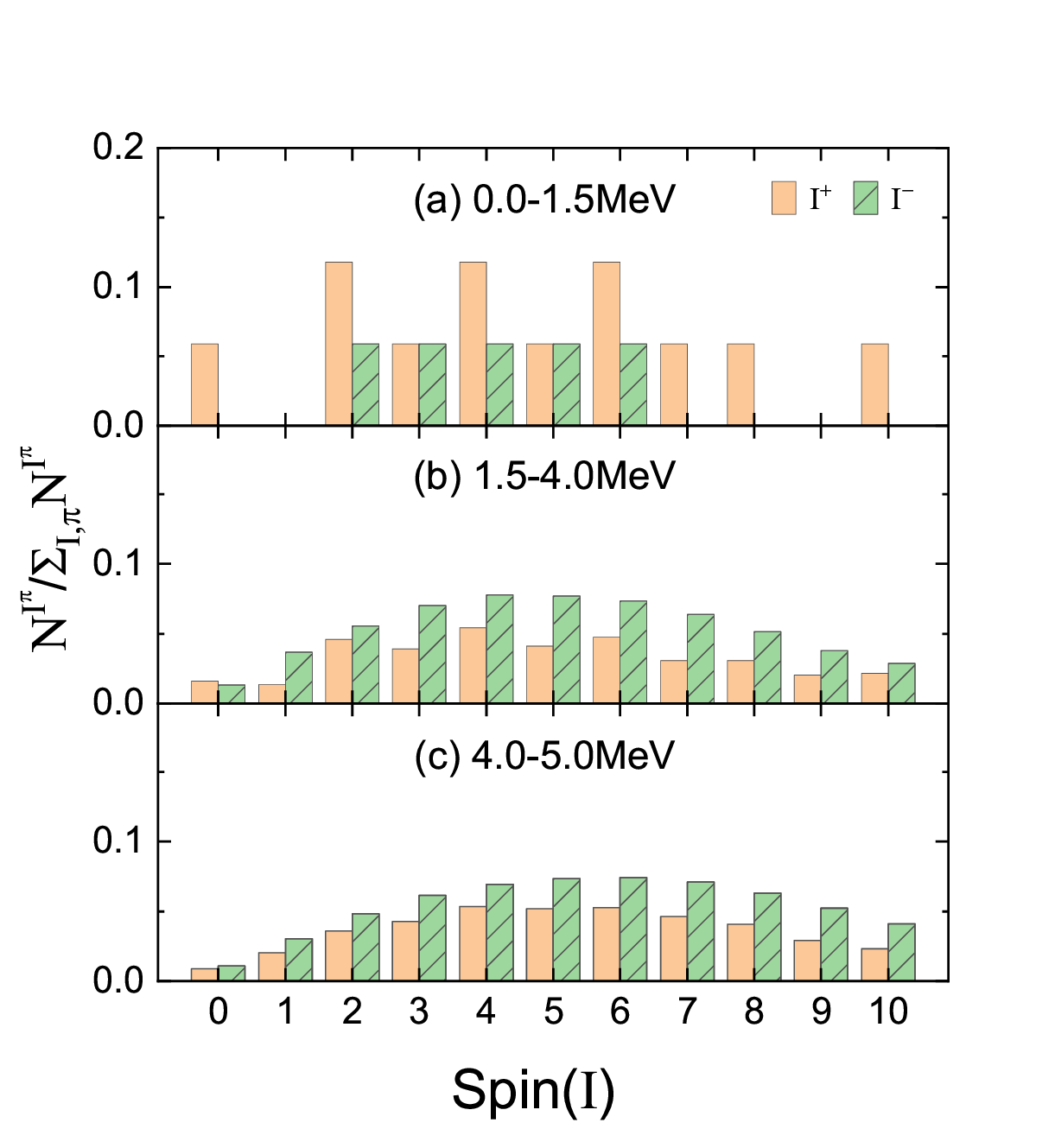}
    \caption{Normalised spin distribution of nuclear levels in $^{164}$Dy for positive and negative parities in three different excitation energy regimes, namely: (a) collective, (b) pair-breaking, and (c) multi-qp or chaotic regimes. The Figure is taken from  Ref.~\cite{jiaqi2023}.}
    \label{fig:spin-new.eps}
\end{figure}
\section{Results and discussions}
Figure~\ref{fig:NLD_compare} represents the density of levels in $^{164}$Dy from our shell model calculations, which are compared to the spectroscopic measurements as well as the data provided by the Oslo experiment. Our calculation generates nuclear states individually each having definite spin and parity, which allows a quantitative comparison with the discrete levels obtained from spectroscopic measurements feasible. %From such a comparison, we have found that our model well reproduces experimentally observed discrete nuclear levels at low energies  (up to which the data are complete). 
To draw Figure~\ref{fig:NLD_compare}, we have derived the density of individual nuclear levels using the concept of energy bins. We have divided the entire excitation energy range into bins of 120 keV widths and counted all the levels within each energy bin. Therefore, in Figure~\ref{fig:NLD_compare}, in the NLD curve of our PSM calculation as well as that of discrete level data, there are about eight data points within 1 MeV which is consistent with the $^{164}$Dy Oslo NLD curve \cite{Oslo1}. %For known discrete levels in $^{164}$Dy with experimentally assigned spin, except for the first two collective bands, namely, the ground-state band and the $2^+$ $\gamma$-band starting from 762 keV, where the measurement was extended to high-spin states, spin quantum numbers seldom exceed 10$\hbar$. 
 In Figure~\ref{fig:NLD_compare}, %in which we have for comparison of our calculated NLDs with the known discrete ones,
we have included the calculated levels with spins $I\le 10\hbar$ only. This is justified since in the observed discrete levels in $^{164}$Dy, experimentally assigned spin values are mostly within 10$\hbar$ (except for the first two collective bands, namely, the ground-state band and the $2^+$ $\gamma$-band starting from 762 keV, in which  measurement was extended to high-spins) \cite{NNDC_database}. We have also restricted the excitation energy range up to 4.5 MeV. States within this energy range contain 0-qp, 2-qp, 4-qp, and some 6-qp configurations.  
\begin{figure*} \centering
\includegraphics[scale=0.76]{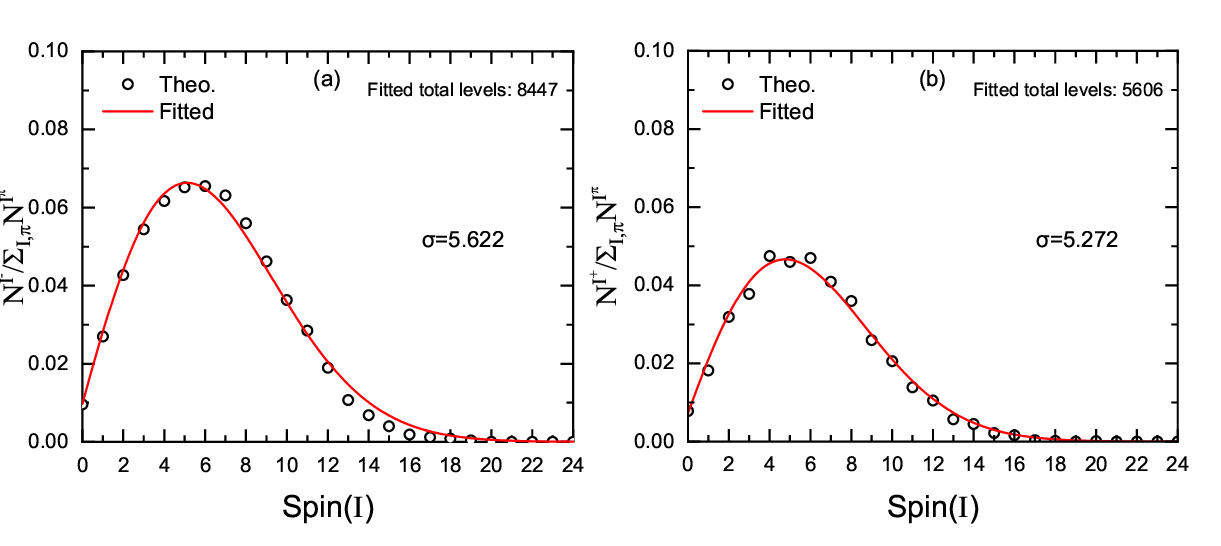}
\caption{\label{fig:gaussians}(Color online) 
Least-squares fitting with Ericson's formula of Eq.~(\ref{spin-distribution}) for the spin-distribution in the levels within $4.0-5.0$ MeV, for (a) odd-parity, and (b) even-parity. The Figure is taken from Ref.~\cite{jiaqi2023}.}
\end{figure*}
According to the underlying mechanism of level formation with the increasing excitation, we divide the entire NLD curve of Figure~\ref{fig:NLD_compare} into three regimes, namely: (a) collective regime (0 - 1.5 MeV), (b) pair-breaking regime (1.5 - 4.0 MeV), and (c) multi-qp or chaotic regime (beyond 4.0 MeV and above), and briefly discuss the characteristic features in each regime. %The levels of the collective regime, the nuclear structures are well-known from the discrete spectroscopic measurements. %This excitation energy regime contains some pair-breaking configurations in addition to the collective rotational and vibrational excitations. 
It can be seen that nearly perfect agreement between our calculation and the NLD of measured discrete levels up to 1.75 MeV  (i.e., up to which the low-energy discrete level data may be complete) has been attained, including the two up-rises, the first one at 0.75 MeV and the second one at 1.75 MeV. The first up-rise is due to collective vibrations while the second one is now confirmed from our calculation to be due to the onset of pair-breaking following which broken-pair multi-qp configurations gradually contribute. This undoubtedly proves the power of our model in predicting collective as well as pair-breaking states and at the same time, justifies its applicability to the regions of excitation which are out of reach of the present-day experiments. NLD in the pair-breaking regime (where interplay between the collective and the pair-breaking configurations takes place \cite{Wang2020}), has important structure consequences. In particular, here, our curve shows a step-like structure ranging over 2.3 to 4.0 MeV (including a plateau region from 2.4 to 2.9 MeV where the NLD remains nearly constant). This is the energy region where any structure effect in level density and $\gamma$ strengths may sensitively influence neutron capture cross-sections \cite{Mumpower2023}. To our knowledge, this structure is unknown in any previous theoretical and experimental NLD curves including the Oslo one. We also conclude that the third up-rise that can be seen to start at 3.0 MeV bin is a result of concurrent breaking of two nucleon-pairs: a neutron- and a proton-pair and therefore, creation of the first 4-qp state. The energy required to break a neutron-pair and a proton-pair is $2\Delta_n$ and $2\Delta_p$, respectively, and therefore, the lowest 4-qp state occurs at $2(\Delta_n+\Delta_p)$, which is about 3 MeV in $^{164}$Dy. As eventually many such 4-qp configurations and later 6-qp configurations participate in level formation, this rise can be seen in Figure~\ref{fig:dy164-piecewise-multiqp} to continue further resulting in an exponential trend. 

Inclusion of higher-spin states and/or of higher excitations in Figure \ref{fig:NLD_compare} is possible. However, with further extension of our configuration space, computation becomes time-consuming. Besides, NLD beyond 4.5 MeV does not necessarily require precise spectroscopic properties since with increasing density of levels, the curve attains the so-called `stochastization', a term coined by Zelevinsky {\it et al.} \cite{sd-shell_full_diag_1} and therefore, structural details for individual levels no longer remains important. 
For the energies $E>4.5$ MeV, if we do not burden ourselves with exact calculation, i.e., if we give up the time-consuming diagonalization procedure, we can draw an NLD curve up to a much higher energy with the levels obtained without conducting the configuration mixing calculation. The so-drawn curve is qualitatively similar to Figure \ref{fig:NLD_compare} whereas obtained with relatively greatly reduced numerical cost. Figure~\ref{fig:dy164-piecewise-multiqp} shows the result of such calculation for $^{164}$Dy where we have included all projected 0-, 2-, 4-, and 6-qp states of Eq.~(\ref{shell_config}). Additionally, the junction energies from where multi-qp state of a definite seniority enters into the calculation can also be clearly visible. It can also be seen that the calculation nicely reproduces the NLD data at the neutron separation energy (= 7.658 MeV)  estimated using the measured $s$-wave neutron resonance level spacing.
%Therefore, our finding in the pair-breaking region revises the usual consideration that the exponential behavior of the NLD sets in immediately when the first nucleon pairs are broken at $E> 2\Delta$ \cite{Guttormsen2015} (which implicitly assumed the same $\Delta$ for neutrons and protons). Instead, our calculation suggests a much-delayed observation of the exponential behavior in NLD, due to the recognition of the simultaneous breaking of a neutron- and a proton-pair. %Another conclusion is that the discrete of the discrete NLD curve starting from 2.0 MeV energy bin therefore implies that a large amount of 2-qp and almost all the 4-qp configurations are missing in the spectroscopic measurement. %This implies that structure dominance by individual configurations becomes more pronounced at the collective and pair-breaking regimes. 

Spin distribution in NLDs is another topic under intensive discussion \cite{Grimes2016}. Figure~\ref{fig:spin-new.eps} shows the distribution of NLD for separate parities (normalised to total NLD) as a function of angular momenta. In general, a Gaussian distribution is considered for spins following the early work of T. Ericson \cite{Ericson1959,Ericson1960} which was based on a statistical model that assumed random coupling of angular momenta:
\begin{equation}\label{spin-distribution}
\rho(E, I) \approx \rho(E) \frac{2I+1}{2{\sqrt{2\pi}}\sigma^3} \exp{-\frac{I(I+1)}{2\sigma^2}},
\end{equation}
where $\rho(E)$ is the spin-independent level density and $\sigma$, known as the dispersion, is an unknown parameter closely linked to nuclear structure and is responsible for determining the shape of the Gaussian. In our work, Eq.~(\ref{spin-distribution}) seems to be applicable only in the chaotic regime. %beyond a certain excitation energy ($\sim$4 MeV in the present example of $^{164}$Dy.
In other words, our finding suggests clear structure-dependence of spin distribution at different excitation energy regimes. Figure~\ref{fig:spin-new.eps}(a) represents the spin distribution in the collective regime in which only 17 known levels exist among which only 5 are of odd parity and therefore, obviously, irregular spin distributions can be seen for both the parities. The middle plot, Figure~\ref{fig:spin-new.eps}(b), shows the spin-parity dependence of the pair-breaking regime %which acts as the transitional regime between the collective and chaotic ones and which is affluent with interesting structural details. This is the energy regime throughout which the breaking of all nucleon pairs takes place. 
where more number of odd-parity states than even-parity ones can be visible. Noticeably, for the even-parity states, there is an odd-even staggering effect %in which an even-spin distribution can be seen as larger than either of its neighboring odd-spin ones.
which, however, is not the case with the odd-spin distribution that tends to take the shape of Gaussian. Finally, in Figure~\ref{fig:spin-new.eps}(c), as the states enter the chaotic regime, consistent with Ericson's finding, perfect Gaussian distribution emerges for both parities which therefore implies that application of Eq.~(\ref{spin-distribution}) makes sense in this regime of nuclear excitation. Ericson suggested a relation for determining the value of the dispersion $\sigma$ from the rigid moment of inertia approximation \cite{Ericson1960} which, however, is not the reality at low excitations as moment of inertia is a rapidly changing quantity there. We have therefore further done a numerical fitting of our spin distribution in the chaotic regime with that of Eq.~(\ref{spin-distribution}) to find the value of $\sigma$ from our shell model calculations. To see the long-range tail in the high-spin part of the Gaussian, we include levels with spins up to $I=24\hbar$ and we obtain $\sigma=5.622$ and $5.272$ for positive- and negative-parity, respectively.  

\section{Summary}\label{conclusion}
By microscopically solving nuclear many-body eigenvalue problem using our novel projected shell model, we have found a one-to-one correspondence of nuclear states with the data from spectroscopic measurements. %Our study nicely explains that the first oscillatory step structure in NLD curve is due to nuclear collective vibrations %(where all nucleons remain coupled as Cooper pairs in nuclear orbitals) while the second one is due to beginning of pair phase transition. %(from when gradual quenching of pair correlations in Cooper condensates starts). %Using quantitative details, we have shown that the distinct oscillatory structures in the low-energy NLD curve the collective excitation and nucleon-pair breaking, the exponential growth of levels in the higher-energy NLD can be described by the combination of the broken-pair states. 
%Our quantitative analysis nicely explains the distinct oscillatory features in the NLD curve at different excitation energy regimes. 
The unprecedented prediction, i.e., the third step structure from $\sim$ 2.3 to 4.0 MeV of excitation energy suggests a need to amend the common belief that the NLD curve starts to grow exponentially as soon as the first Cooper pair breaks apart \cite{Guttormsen2015}. Instead, our study suggests a delayed start in exponential rise since we have seen that a proton-pair as well as a neutron-pair break apart simultaneously, and 4-qp states appear in the system from the combination of those broken pairs. This is a unique realization in fermionic nuclear systems because of the presence of the isospin degree of freedom and hence, of two kinds of pairing gaps. We have also nicely demonstrated that the exponential growth of NLD can be described by the combination of broken-pair multi-qp states. Moreover, our study on the spin distribution of NLD implies prominent structure characteristics different from the mainstream notion based on statistical assumptions that do not take into account shell effects.  %In brief, we have found that different from the usual notion, for low-energy regions before the consummation of pair breaking, spin distribution patterns are generally irregular.  Eventually, as energy increases and reaches 4 MeV and beyond, a bell-shaped distribution appears, consistent with that of the statistical model assuming random coupling of angular momenta. 

\section{Acknowledgments}
Valuable discussions with M. Wiedeking and A. V. Voinov are gratefully acknowledged. This work is supported by the National Natural Science Foundation of China (Grant Nos. 12235003, 12275225, and U1932206).

\end{document}